\documentclass{PoS}

\usepackage{booktabs}
\usepackage{xspace}
\usepackage{multirow}

\usepackage[
  tbtags
]{amsmath}
\usepackage[
  product-units = single,
  output-product = \times,
  inter-unit-product = \cdot
]{siunitx}

\newcommand{\eV}{\ensuremath{\text{e}\mspace{-0.8mu}\text{V}}\xspace}

\newcommand{\GeV}{\text{G\eV}\xspace}
\newcommand{\TeV}{\text{T\eV}\xspace}

\newcommand{\gambit}{\textsf{GAMBIT}\xspace}

\newcommand{\GB}{\gambit}

\newcommand{\buckfast}{\textsf{BuckFast}\xspace}

\newcommand{\pythia}{\textsf{Pythia}\xspace}

\newcommand{\higgsbounds}{\textsf{HiggsBounds}\xspace}

\newcommand{\higgssignals}{\textsf{HiggsSignals}\xspace}
\newcommand{\ds}{\textsf{DarkSUSY}\xspace}
\newcommand{\darksusy}{\ds}

\newcommand{\micromegas}{\textsf{micrOMEGAs}\xspace}

\newcommand\FlexibleSUSY{\textsf{FlexibleSUSY}\xspace}

\newcommand\HDECAY{\textsf{HDECAY}\xspace}

\newcommand\SDECAY{\textsf{SDECAY}\xspace}
\newcommand\SUSYHIT{\textsf{SUSY-HIT}\xspace}

\newcommand\susyhit{\SUSYHIT}
\newcommand\gmtwocalc{\textsf{GM2Calc}\xspace}

\newcommand\superiso{\textsf{SuperIso}\xspace}

\newcommand\nulike{\textsf{nulike}\xspace}

\newcommand\MultiNest{\textsf{MultiNest}\xspace}
\newcommand\multinest{\MultiNest}

\newcommand\diver{\textsf{Diver}\xspace}
\newcommand\ddcalc{\textsf{DDCalc}\xspace}

\title{First SUSY results with GAMBIT}

\ShortTitle{First SUSY results with GAMBIT}

\author{\speaker{Anders Kvellestad} \normalfont{on behalf of the GAMBIT collaboration}\\%
       NORDITA, Roslagstullsbacken 23, SE-10691 Stockholm, Sweden\\
       E-mail: \email{anders.kvellestad@nordita.org}}

\abstract{
Based on arXiv:1705.07935 and arXiv:1705.07917, we present the first global fits of supersymmetric models using the new Global And Modular Beyond-the-Standard-Model Inference Tool (\GB). With \GB we have performed frequentist fits of the GUT-scale CMSSM, NUHM1 and NUHM2 models, as well as the weak-scale MSSM7, extending existing results in terms of the number of observables included, scanning techniques and treatment of nuisance parameters. For the GUT-scale models our analyses show that a stop co-annihilation scenario provides the best fit to current data, and that in the CMSSM the stau co-annihilation scenario is ruled out at 95\% confidence level. For the MSSM7 we find that the best-fit scenario has light higgsinos and highly under-abundant relic density due to efficient chargino co-annihilation.
}

\FullConference{The European Physical Society Conference on High Energy Physics\\
         5-12 July\\
         Venice, Italy}

\begin{document}

\section{Introduction}

Supersymmetry (SUSY) has long been among the most promising avenues for new physics, potentially offering solutions to many problems left unresolved in the Standard Model (SM).  The phenomenological richness of SUSY also implies that the parameter space of SUSY theories can be constrained by a wide range of experimental searches.  To investigate the combined impact of current results from all relevant experiments requires a `global fit' -- a comprehensive and statistically rigorous comparison between theory predictions and data across the parameter space probed.

For this purpose we have developed \GB, a new open-source tool for performing large-scale global fits of Beyond-the-Standard-Model (BSM) theories \cite{gambit,SDPBit,DarkBit,ColliderBit,FlavBit,ScannerBit}.  While we here present the SUSY global fits we carried out in \cite{CMSSM,MSSM}, \GB itself is designed as a theory-agnostic tool capable of performing fits of any BSM theory where the necessary tools for calculating theory predictions exist.  For an example of a non-SUSY analysis with GAMBIT we refer to our global fit of the scalar singlet model in \cite{SSDM}.

We have performed global fits of four different models within the larger framework of the Minimal Supersymmetric Standard Model (MSSM): the Constrained MSSM (CMSSM) and its Non-Universal Higgs Mass extensions (NUHM1 and NUHM2), with parameters defined at the GUT scale, and a 7-parameter version of the MSSM (MSSM7) defined at a scale of $Q=1$~\TeV.  
The free parameters of each model, along with the ranges and sampling priors used in our parameter scans, are provided in Table 1 of Ref.\ \cite{CMSSM} (CMSSM, NUHM1 and NUHM2) and Table 1 of Ref.\ \cite{MSSM} (MSSM7). 
In our scans we have also varied five nuisance parameters: the SM strong coupling ($\alpha_s$), the top mass ($m_t$), the local dark matter (DM) density ($\rho_0$) and two nuclear matrix elements ($\sigma_s$ and $\sigma_l$) relevant for DM direct detection.  To ensure a thorough exploration of each model's parameter space we have performed a number of scans with varying priors, sampling algorithms and settings. The final results are based on the combined set of parameter samples for each model.  The scans were performed using the differential evolution scanner \diver~\textsf{1.0.0} \cite{ScannerBit}
and the nested sampler \multinest~\textsf{3.10} \cite{MultiNest}.  We refer to \cite{CMSSM,MSSM} for further details on the scanning procedure.

The likelihood function in our analyses includes contributions from Higgs measurements; electroweak precision observables; a large number of flavour observables; direct SUSY searches at LEP and LHC (Runs I and II); the DM relic density; and multiple direct and indirect DM searches. As a conservative choice we have implemented the observed relic density only as an upper limit, thus not penalizing parameter regions that predict an under-abundant relic density.

For calculations of the theory predictions we have made use of the following tools: \FlexibleSUSY~\textsf{1.5.1}~\cite{Athron:2014yba}, \HDECAY~\cite{Djouadi:1997yw} and \SDECAY~\cite{Muhlleitner:2003vg} through \susyhit~\textsf{1.5}~\cite{Djouadi:2006bz}, \higgsbounds~\textsf{4.3.1}~\cite{Bechtle:2008jh,Bechtle:2011sb,Bechtle:2013wla}, \higgssignals \textsf{1.4.0}~\cite{HiggsSignals}, \gmtwocalc~\textsf{1.3.1}~\cite{gm2calc}, \superiso~\textsf{3.6}~\cite{Mahmoudi:2007vz,Mahmoudi:2008tp}, \pythia~\textsf{8.212}~\cite{Sjostrand:2006za,Sjostrand:2014zea}, \buckfast~\cite{ColliderBit}, \micromegas~\textsf{3.6.9.2}~\cite{micromegas}, \darksusy~\textsf{5.1.3}~\cite{darksusy}, \ddcalc~\textsf{1.0.0}~\cite{DarkBit} and \nulike~\textsf{1.0.4}~\cite{IC22Methods,IC79_SUSY}.

The \GB input files, generated likelihood samples and best-fit benchmark points for the analyses presented here are available online through \textsf{Zenodo} \cite{the_gambit_collaboration_2017_801642,the_gambit_collaboration_2017_801640}.

\section{Results}

\begin{figure*}
  \centering
  \includegraphics[width=0.49\textwidth]{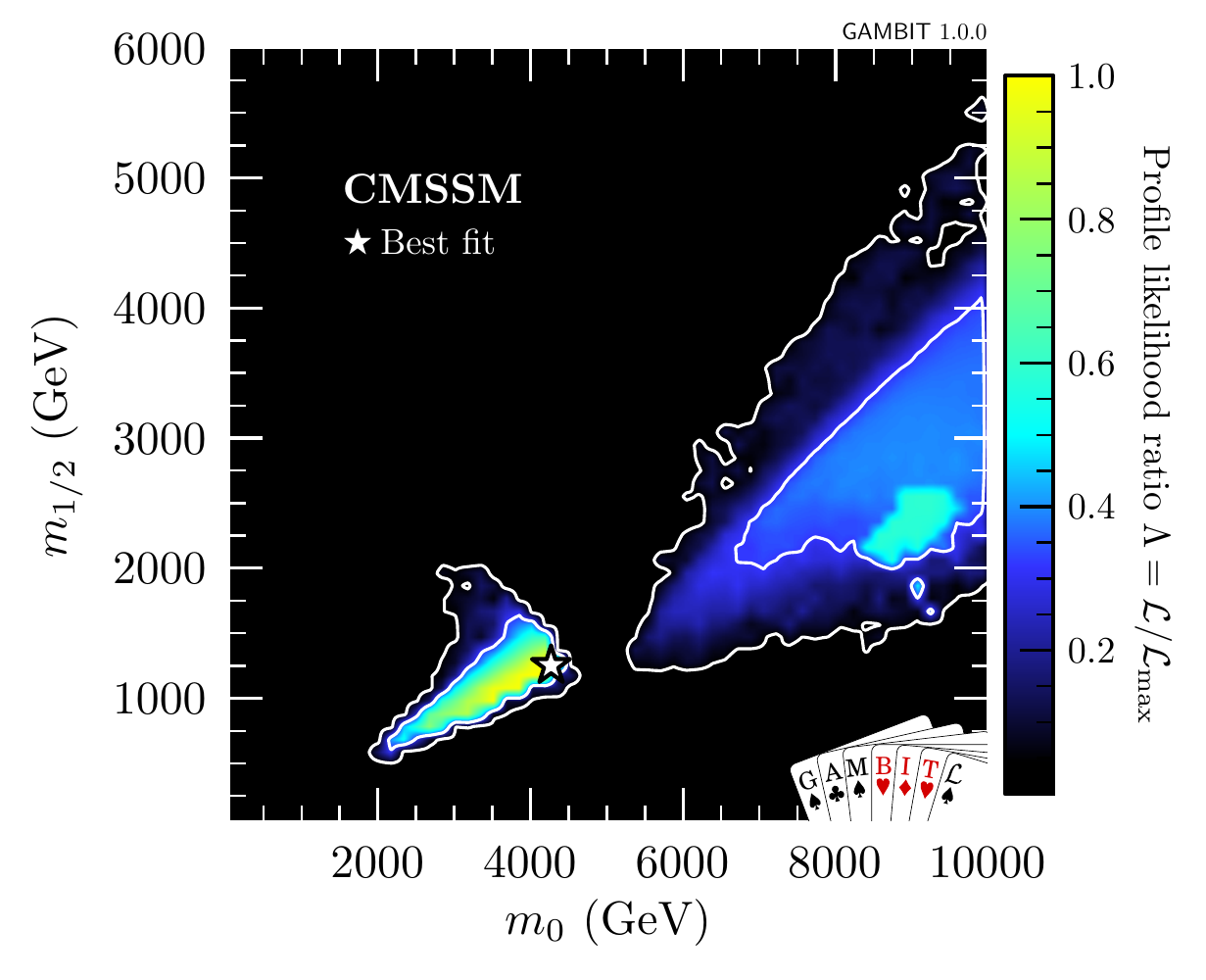}
  \includegraphics[width=0.49\textwidth]{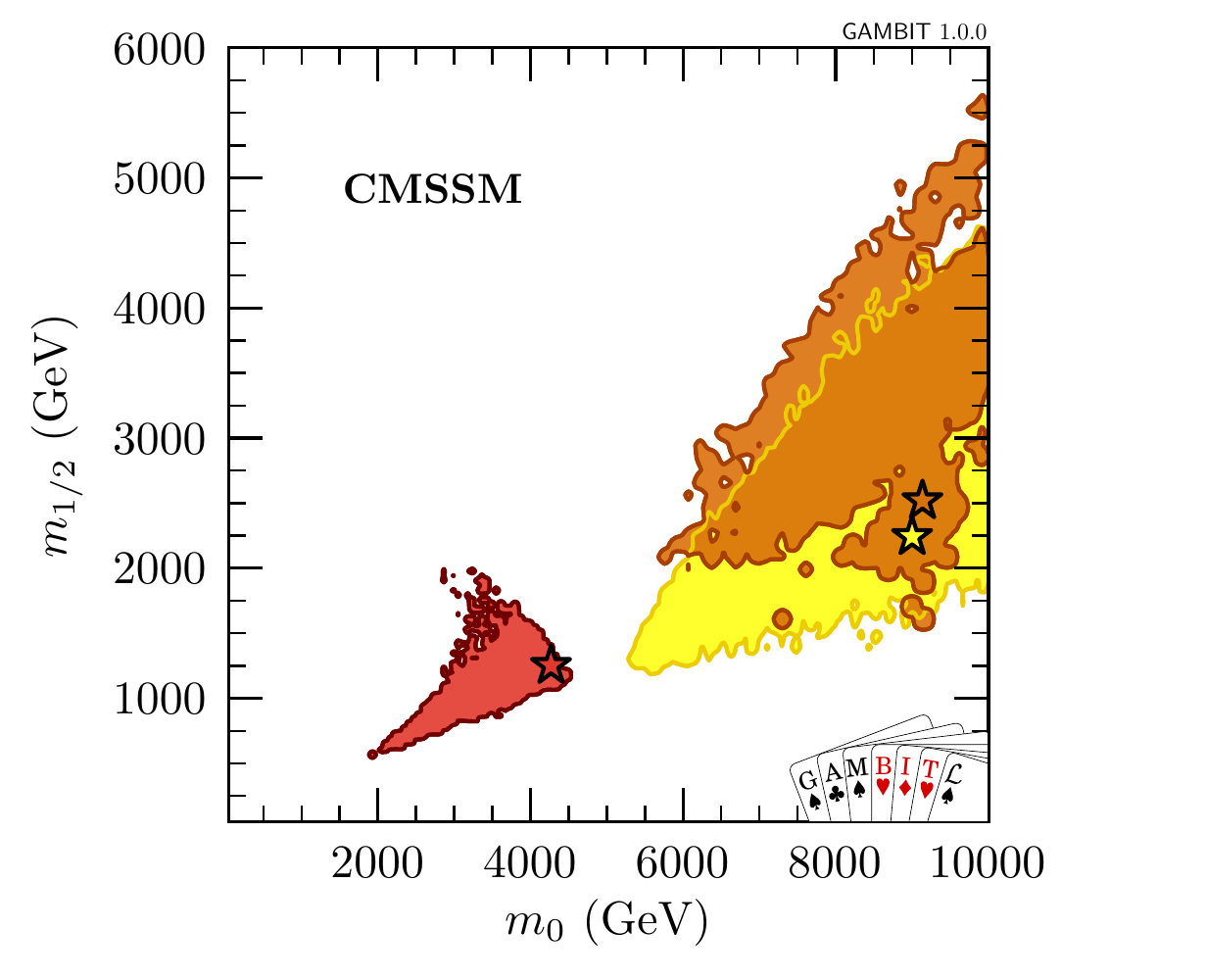}\\
  \includegraphics[width=0.49\textwidth]{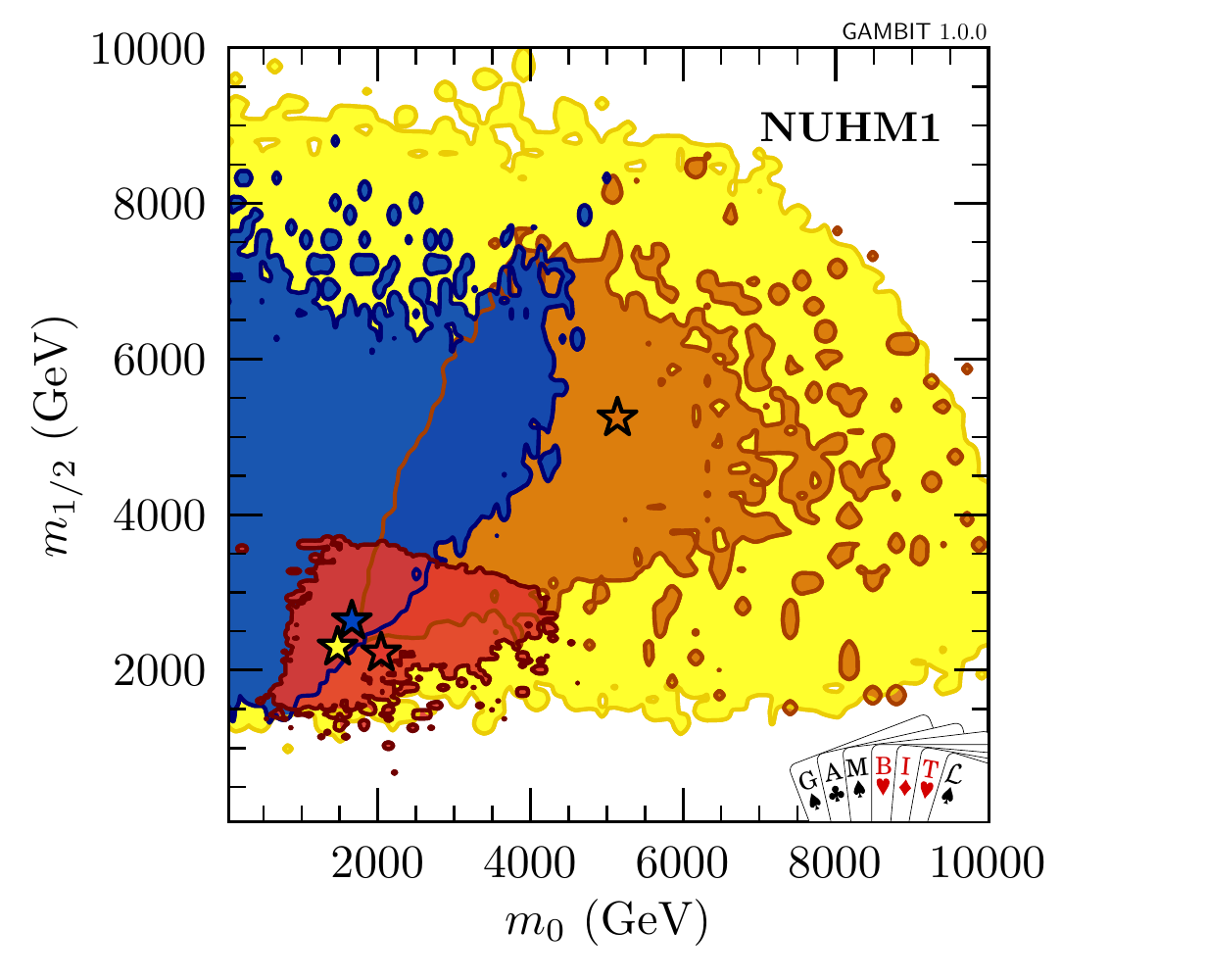}
  \includegraphics[width=0.49\textwidth]{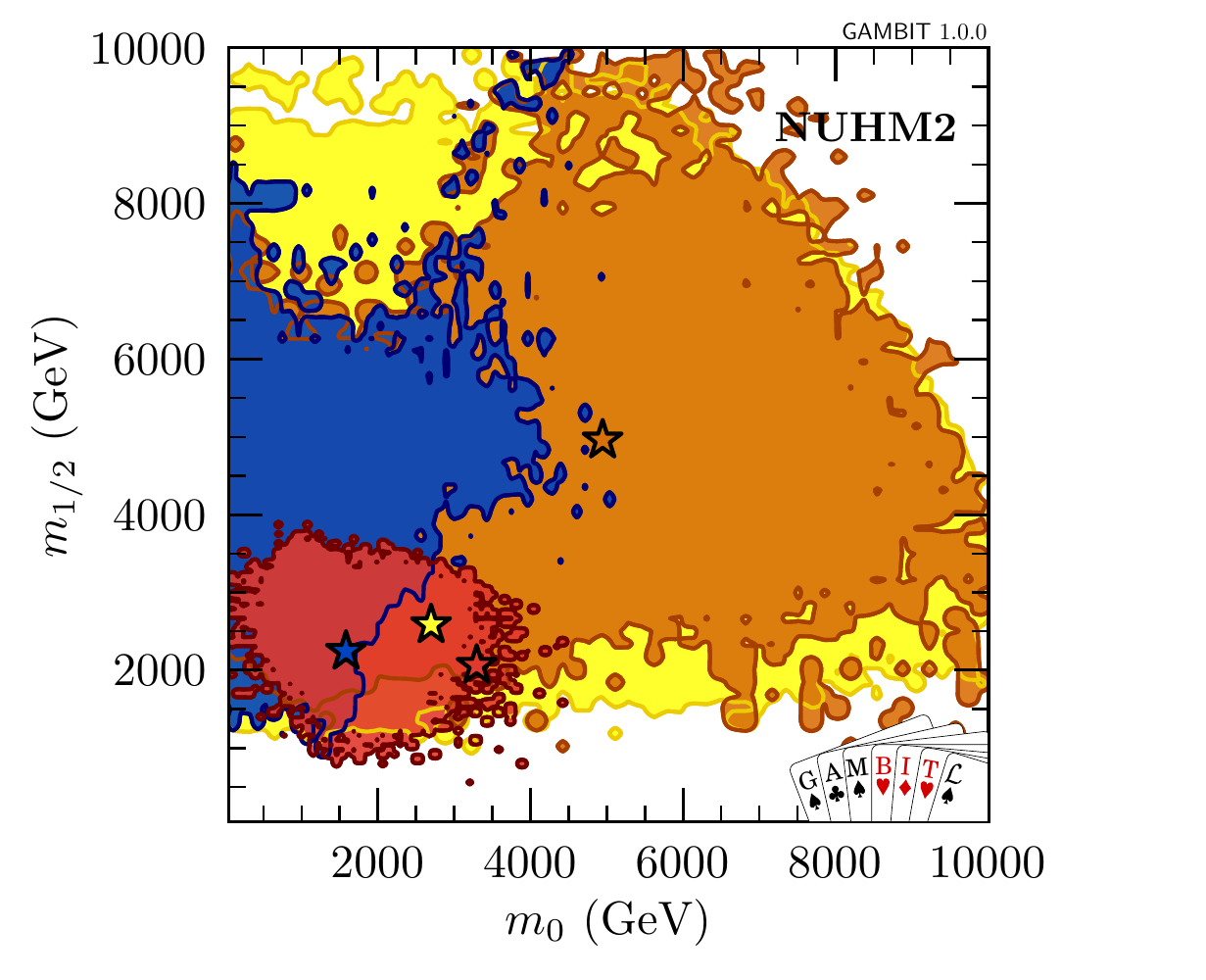}
  \includegraphics[height=4mm]{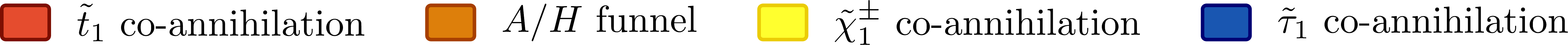}
  \caption{
  \textit{Top left:} The profile likelihood ratio in the $m_0$-$m_{1/2}$ plane of the CMSSM. The white lines depict the 68\% and 95\% CL contours while the white star indicates the best-fit point. 
  \textit{Top right:} Colouring of the 95\% CL region to indicate which mechanisms contribute to keeping the neutralino relic density below the observed value. Note that the colouring is not exclusive, \textit{i.e.}\ overlapping colours indicates that multiple mechanisms may contribute in the given region.
  \textit{Bottom:} Similarly coloured plots for the $m_0$-$m_{1/2}$ planes of the NUHM1 (left) and NUHM2 (right). Figures from~\cite{CMSSM}.
  }
  \label{fig:m0_m12_cmssm_nuhm}
\end{figure*}

\begin{figure*}
  \centering
  \includegraphics[width=0.49\textwidth]{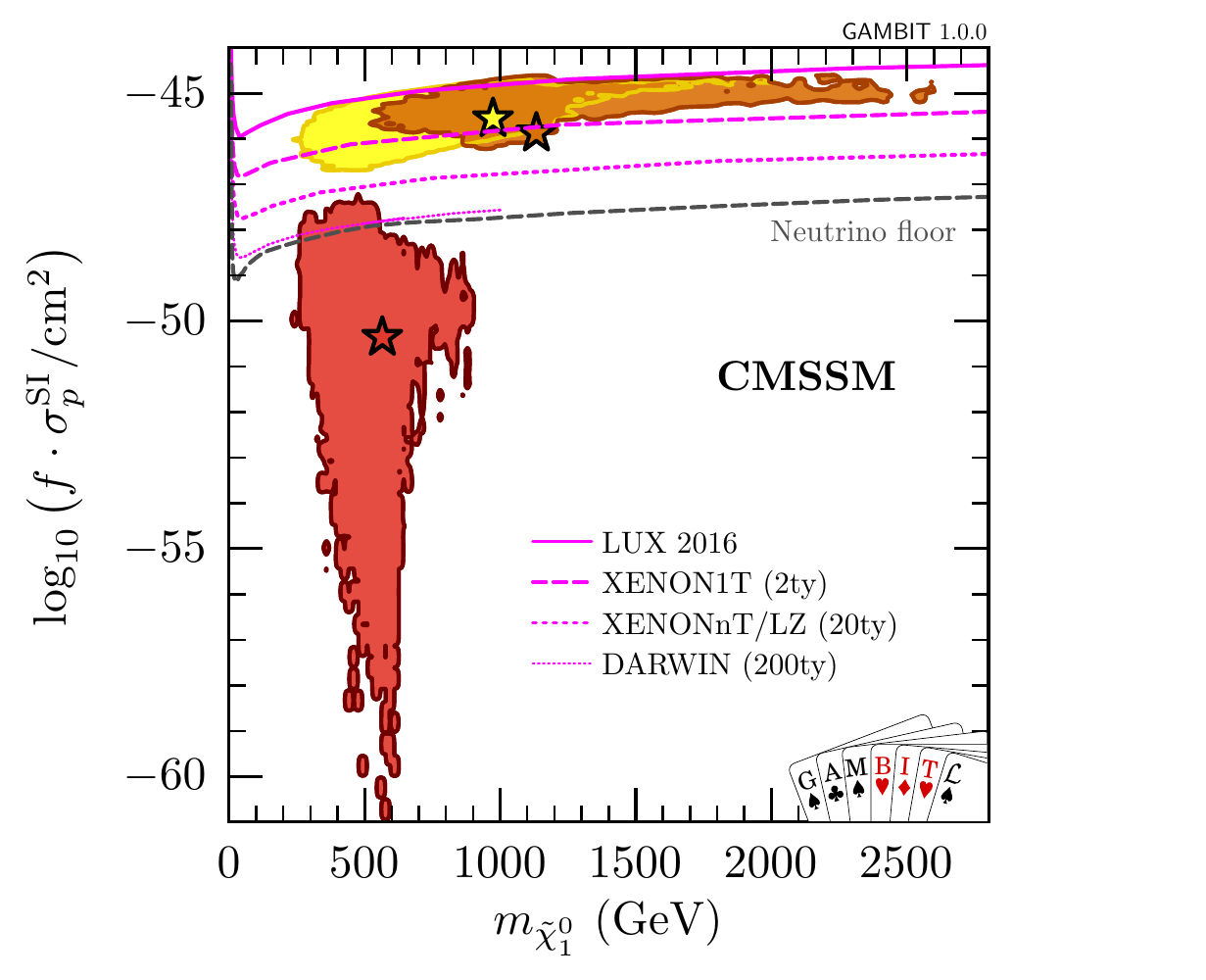}
  \includegraphics[width=0.49\textwidth]{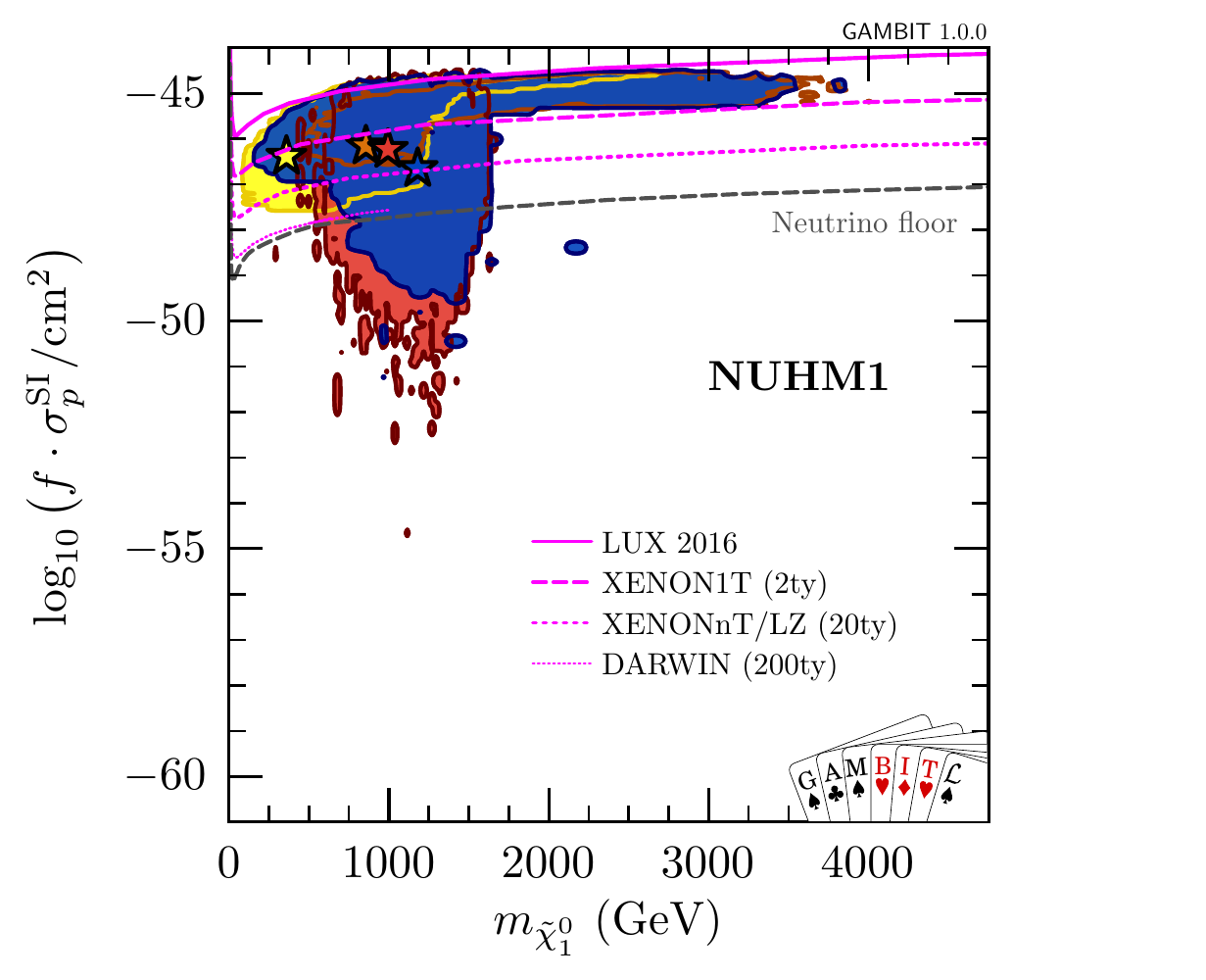}
  \includegraphics[height=4mm]{figures/rdcolours4.pdf}
  \caption{
  \textit{Left:} The CMSSM 95\% CL region in the plane of the neutralino mass and the spin-independent neutralino-proton scattering cross-section. The final 90\% CL exclusion limit from LUX \cite{LUXrun2} along with prospective limits from future direct detection searches \cite{XENONnTLZ,DARWIN} are displayed as pink lines.
  \textit{Right:} A similar plot for the NUHM1. Figures from~\cite{CMSSM}.
  }
  \label{fig:mChi_sigmaSI_cmssm_nuhm}
\end{figure*}

Figure~\ref{fig:m0_m12_cmssm_nuhm} (top row) shows the results in the $m_0$-$m_{1/2}$ plane of the CMSSM. We find that the best-fit is provided by the stop co-annihilation region in the lower-left part of the plane.  The preferred parameter space also encompasses a region at larger values for the mass parameters, where an over-abundant relic density is avoided through chargino co-annihilation and/or resonant annihilation through the $A/H$-funnel. Interestingly, the stau-coannihilation region is now excluded at 95\% CL in the CMSSM.

Results for the $m_0$-$m_{1/2}$ planes of the NUHM1 and NUHM2 are shown in the bottom row of Figure~\ref{fig:m0_m12_cmssm_nuhm}. Compared to the CMSSM, the additional parametric freedom in the NUHM1 and NUHM2 leads to a widening of the preferred parameter space. The best-fit region is again the stop co-annihilation region, but now also the stau co-annihilation region is allowed at the 95\% CL. 

The stop co-annihilation region in the CMSSM extends down to stop masses of around $250$~\GeV. Due a stop--neutralino mass difference of less than $50$~\GeV it will be challenging to fully explore the stop co-annihilation scenario at the LHC. However, there is hope that at least the low-mass part of this region can be probed by LHC searches in the near future. For the chargino co-annihilation and $A/H$-funnel regions, DM direct detection experiments are most promising for discovery. As shown in Figure~\ref{fig:mChi_sigmaSI_cmssm_nuhm} (left), future searches with XENON1T, XENONnT, LZ and DARWIN will be able to completely exclude these regions in the CMSSM. We note that the smallest scattering cross-sections found in the stop co-annihilation region are due to fine-tuned cancellations in the tree-level matrix elements, expected not to hold at loop order. 

The situation is rather similar for NUHM1 and NUHM2. Future direct detection experiments will fully probe the current chargino co-annihilation and $A/H$-funnel regions, as shown for NUHM1 in the right panel of Figure~\ref{fig:mChi_sigmaSI_cmssm_nuhm}. However, large parts of the preferred stop and stau co-annihilation parameter space will remain out of reach for both direct and indirect DM searches, as well as for future LHC searches.

\begin{figure*}
  \centering
  \includegraphics[width=0.49\textwidth]{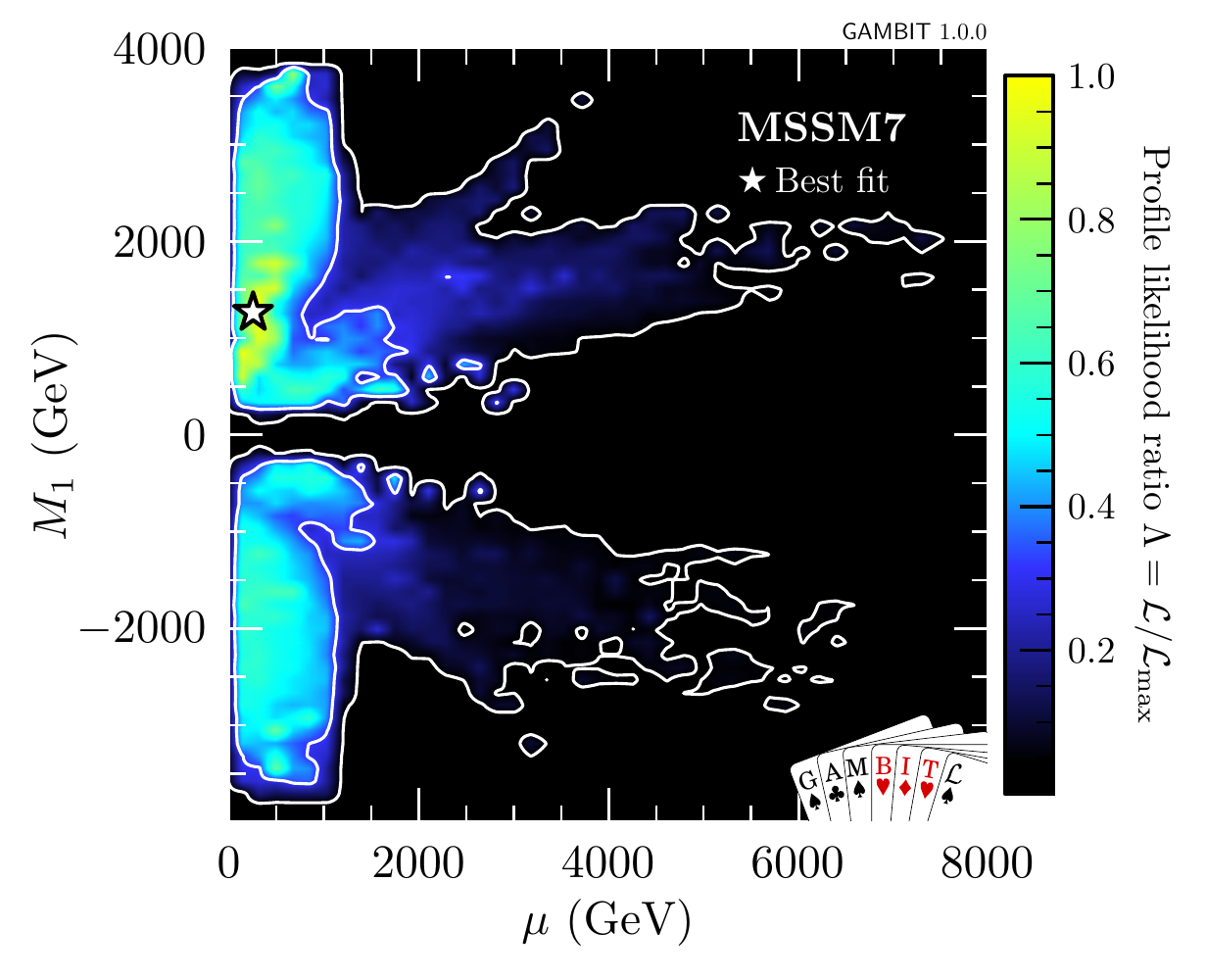}
  \includegraphics[width=0.49\textwidth]{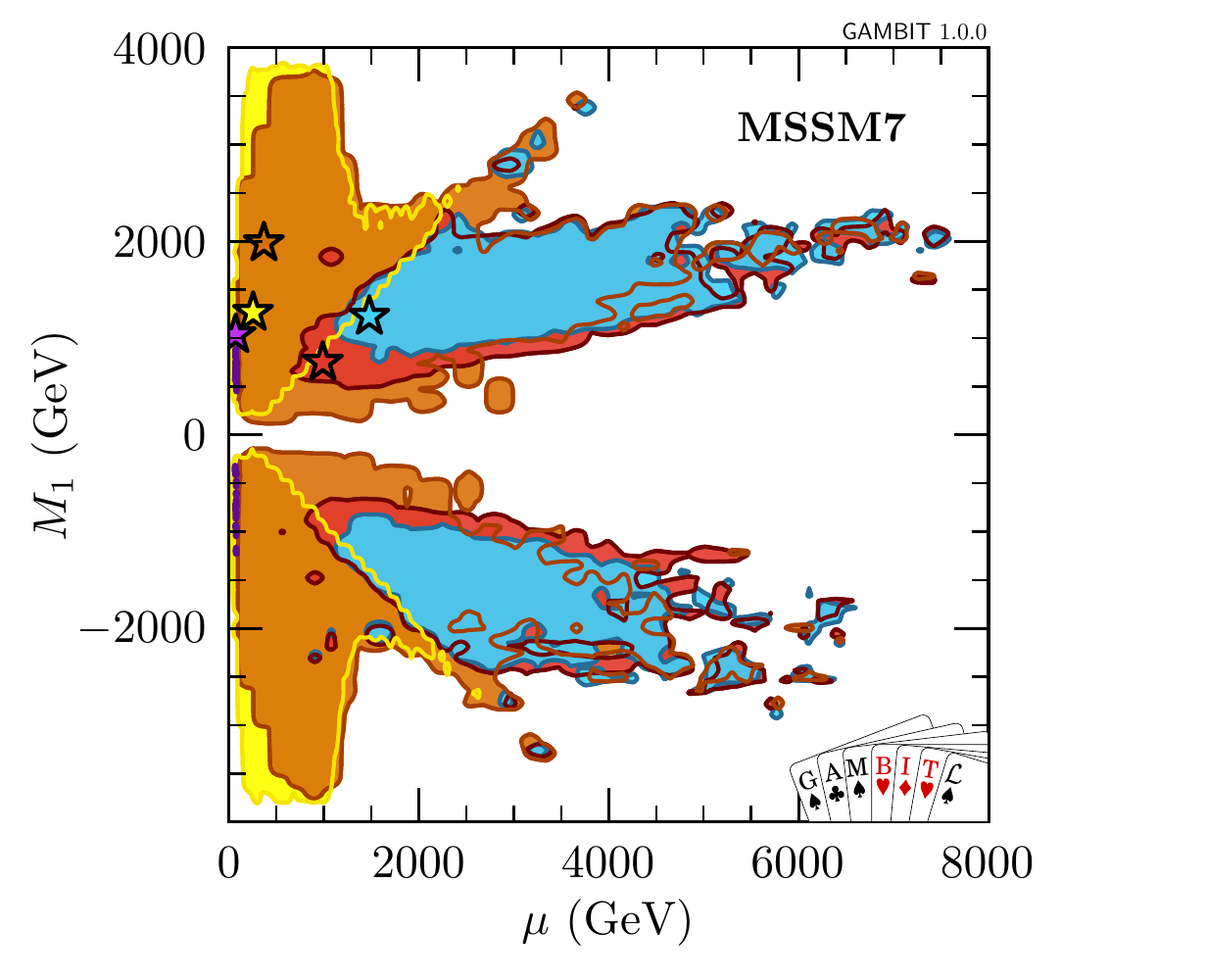}\\
  \includegraphics[width=0.49\textwidth]{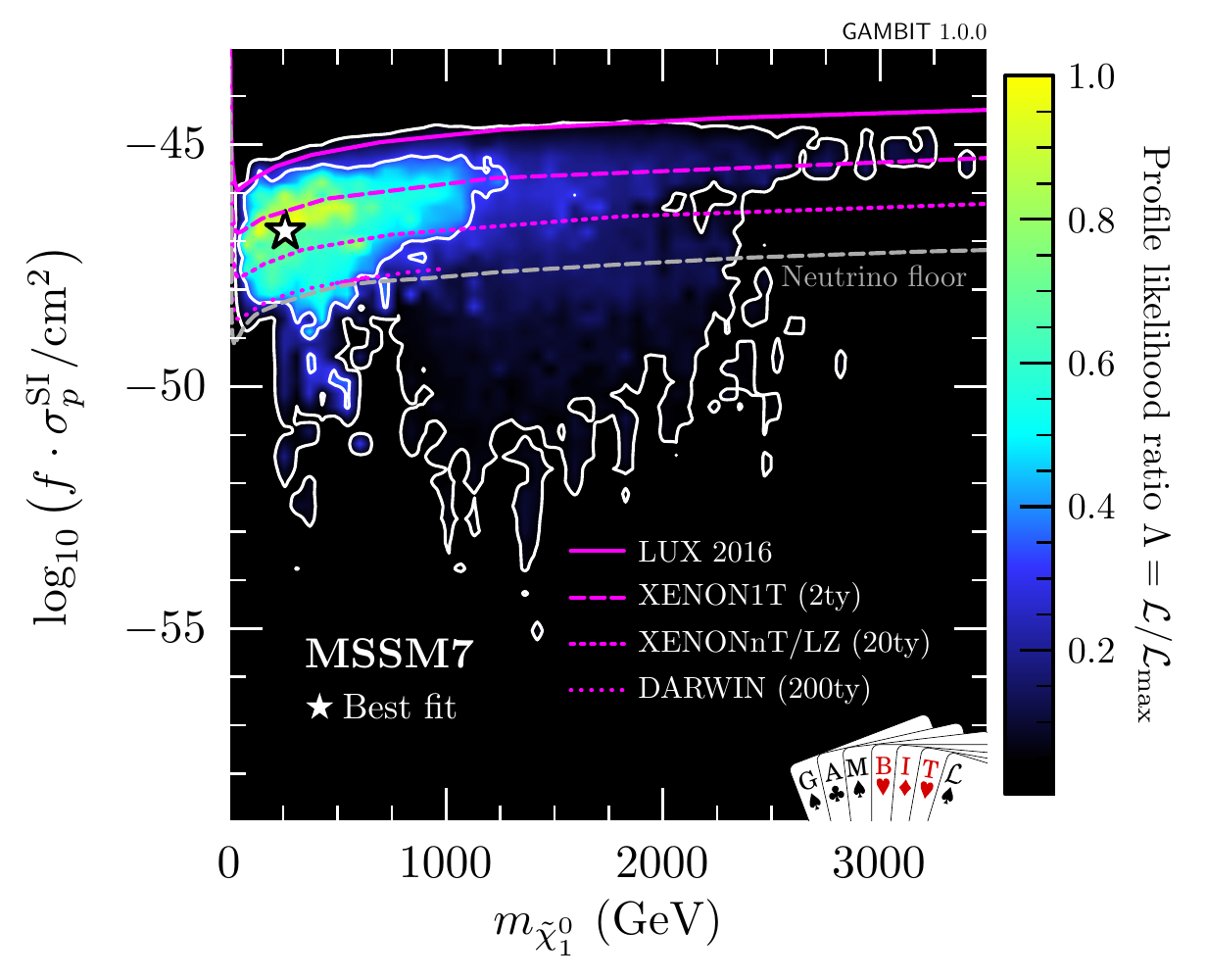}
  \includegraphics[width=0.49\textwidth]{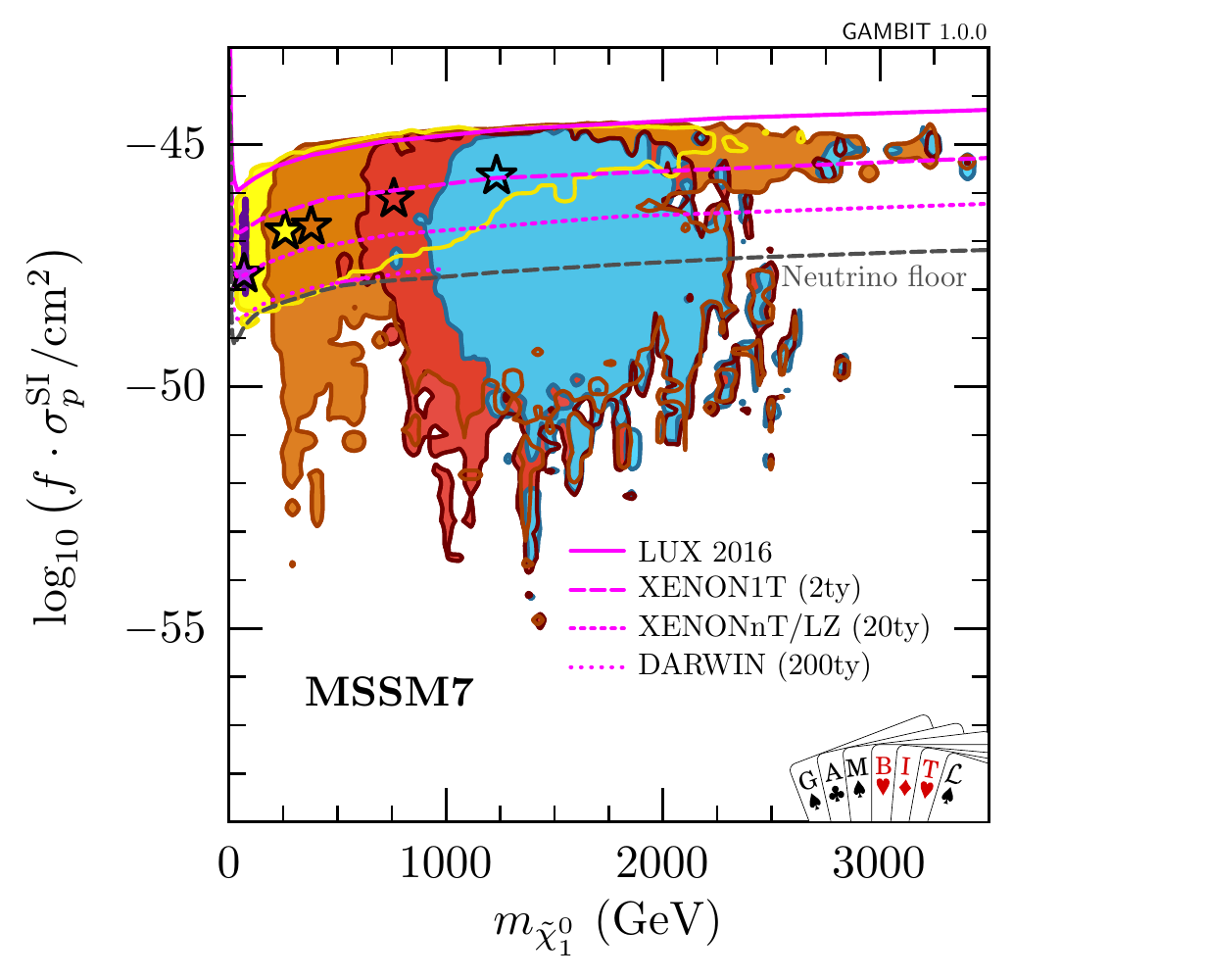}
  \includegraphics[width=0.99\textwidth]{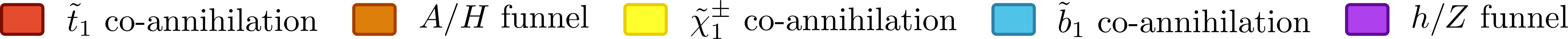}
  \caption{
  \textit{Top left:} The MSSM7 profile likelihood ratio in the plane of the bino mass parameter $M_1$ and the higgsino mass parameter $\mu$. \textit{Top right:} The 95\% CL region in the same coloured according to which mechanisms deplete the relic density. \textit{Bottom:} Similar plots for the plane of the neutralino mass and the spin-independent neutralino-proton scattering cross-section. Figures from~\cite{MSSM}.
  }
  \label{fig:results_mssm7}
\end{figure*}

In the top row of Figure~\ref{fig:results_mssm7} we show the results for the $\mu$-$M_1$ plane of the weak-scale MSSM7. The parameterisation used for the MSSM7 encompasses three possible scenarios for the composition of the lightest neutralino: bino-dominated ($M_1 < \mu$), bino-higgsino mixture ($M_1 \sim \mu$) and higgsino-dominated ($\mu < M_1$). As seen in Figure~\ref{fig:results_mssm7}, all these scenarios are allowed at the 95\% CL. The best-fit parameter space is found in the low-mass part of the chargino co-annihilation region, where the lightest two neutralinos and the lightest chargino are higgsino-dominated with masses around $250$~\GeV. The efficient annihilation of higgsinos leads to a highly under-abundant relic density in this region. 
When the neutralino is bino-dominated, stop and/or sbottom co-annihilation becomes important across large regions of the preferred parameter space. The close correlation between the stop and sbottom co-annihilation regions is due to the MSSM7 being parameterised with a common sfermion soft-mass parameter. The $A/H$-funnel can contribute across almost the entire allowed parameter space. 

As for the GUT-scale models, future DM direct detection searches will fully explore the chargino co-annihilation region. This can be seen in the bottom row of Figure~\ref{fig:results_mssm7}. Also the other relic density mechanisms can be significantly constrained by future direct detection experiments, and to some extent by indirect DM searches (not shown). There is also hope that the low-mass parts of the stop and sbottom co-annihilation regions can be probed at the LHC in the near future.

\section{Conclusion}

We have summarised the comprehensive GAMBIT global fits of the CMSSM, NUHM1, NUHM2 and MSSM7 that we carried out in \cite{CMSSM,MSSM}.  In the CMSSM, NUHM1 and NUHM2 we find that the highest-likelihood parameter regions are found in the stop co-annihilation region, while in the MSSM7 a low-mass chargino co-annihilation scenario provides the best fit to data.  In the CMSSM the stau co-annihilation region is now ruled out at 95\% CL. In all models the currently preferred chargino co-annihilation regions will be fully probed by future direct DM searches, and for the CMSSM, NUHM1 and NUHM2 the same is true for the $A/H$-funnel.

\section*{Acknowledgments}

AK gratefully acknowledges current and former members of the GAMBIT collaboration, with whom this work was carried out.

\bibliographystyle{JHEP_pat}
\bibliography{references_notitles}
\end{document}